\documentclass[aps,prb,two column,showpacs,groupedaddress,amsmath,amssymb]{revtex4-2}
\usepackage{amsmath}
\usepackage{amssymb}
\usepackage[title]{appendix}
\usepackage{graphicx}
\usepackage{bbm}
\usepackage{epsfig}
\usepackage[usenames]{color}
\usepackage{bm}
\usepackage{amsthm}
\usepackage{mathrsfs}
\usepackage{braket}

\usepackage{subfigure}
\usepackage[colorlinks, linkcolor=blue,  anchorcolor=blue, urlcolor=blue, citecolor=blue ]{hyperref}

\newcommand{\nn}{\nonumber}

\newcommand{\be}{\begin{eqnarray}}
\newcommand{\ee}{\end{eqnarray}}

\begin{document}
	
	\title{Quasi-Non-Hermitian Edge Bursts Induced by Nonuniform Loss}
	\author{Ze Yang}
	\author{Wei Li}
	\author{Fuxiang Li}
	\email[Corresponding author: ]{fuxiangli@hnu.edu.cn}
	\affiliation{School of Physics and Electronics, Hunan University, Changsha 410082, China}
	\date{\today}
	\begin{abstract}

    Non-Hermitian quantum walks on lossy lattices with open boundaries can exhibit an anomalous peak of the loss probability at the boundary, known as the non-Hermitian edge burst (NHEB). This phenomenon has been attributed to the combined effect of the non-Hermitian skin effect (NHSE) and a gapless imaginary spectrum. Here we investigate a class of models in which the NHSE is induced by magnetic flux while the loss is spatially nonuniform. We show that the spatial distribution of loss plays a crucial role in determining the emergence and strength of the NHEB. Notably, even in the absence of the NHSE, a weak boundary accumulation of the loss probability persists. We term this effect a quasi–non-Hermitian edge burst (quasi-NHEB). By analyzing the dependence of the loss probability on the initial position, we further demonstrate that the quasi-NHEB obeys a bulk–edge scaling relation distinct from that of conventional NHEB. Our results show that spatially nonuniform loss alone can generate boundary-localized loss anomalies even without the NHSE, providing new insight into non-Hermitian boundary phenomena and a broader platform for their exploration and potential applications.
		
	\end{abstract}

	\maketitle

\section{Introduction}

Non-Hermitian quantum mechanics has attracted sustained attention over the past few decades owing to its unique spectral, dynamical properties and broad relevance to a wide range of physical systems. Intensive research has revealed a variety of unconventional non-Hermitian phenomena, including parity-time (PT) symmetry~\cite{i3,i4,i34,i35,i9review}, non-Hermitian topology~\cite{i24,i25,i26,i27,i28,i29,i31,i40single,i43,i21nhse}, exceptional points (EPs)~\cite{i9op,i10exop,i11op,i13ep}, quantum transport~\cite{i1,i2,i5}, optical realizations~\cite{i7,i6pt,i12ptopex,i14pt,i9op,i10exop,i11op,i13op,i38o,i39o,i41s,i19opt}, and open quantum systems~\cite{i25,i29,i36t,i42s,i37t,i10review}.
Among these research directions, one of the most prominent phenomena is the non-Hermitian skin effect (NHSE), in which the eigenstates exhibit a strong dependence on the boundary conditions~\cite{i15nhse,i16nhse,i17nhse,i18nhse,i20nhse,i21nhse,i22nhse,inh1,inh2,inh3,inh4,inh5,inh6,inh7,inh8,inh9,inh10}. 
Under open boundary conditions (OBC), the NHSE leads to the accumulation of eigenstates near system boundaries, whereas under periodic boundary conditions (PBC) the eigenstates remain extended. 
The understanding of this phenomenon has been greatly advanced by the development of non-Bloch band theory, which introduces the generalized Brillouin zone and provides a consistent framework for describing bulk–boundary correspondence in non-Hermitian systems~\cite{18nonbloch,20correspondence,201dPBCNHSE,24nonblochAmoeba,AGBZ,LW}. 
In one-dimensional systems, the emergence of the NHSE is closely related to the spectral winding number, a topological invariant unique to non-Hermitian physics~\cite{18NHSEwindingnumber}. Beyond its topological classification, controlling the NHSE through external parameters is of both fundamental and practical interest. 
Most early studies achieved such control through asymmetric hopping, which serves as the primary mechanism for inducing the NHSE~\cite{18nonbloch,20correspondence}. 
More recently, increasing attention has been paid to alternative approaches based on on-site potentials or external fields. 
In particular, electric and magnetic fields can effectively generate position-dependent phase factors or potentials that influence the NHSE~\cite{22electricNHSE,21magneticsuppression,24pseudomagneticNHSE,23enhancement2thNHSE}. 
It has been shown that magnetic flux can significantly modify the localization properties of eigenstates, although its overall effect on the NHSE may depend sensitively on system details and parameter regimes~\cite{25controllableNHSE,25magnetic2DNHSE}. 

The NHSE also gives rise to a variety of intriguing boundary phenomena, among which the non-Hermitian edge burst (NHEB) has attracted considerable attention in recent years~\cite{21WangZhong,23nonuniformYuce,24bipolarNHSE,25Yuce,23opensystemNHEB}. The NHEB is characterized by an anomalous accumulation of loss probability near the system boundary during nonunitary dynamics and is generally understood to originate from the interplay between the NHSE and the closing of the imaginary energy gap~\cite{21WangZhong}. Consequently, the conventional NHEB is intrinsically tied to the existence of the NHSE.
Beyond theoretical studies, the NHEB has also been experimentally demonstrated in photonic systems, where the anomalous boundary-localized photon-loss probability can be directly observed~\cite{24NHEBOp,24XiaoEx}.
These developments highlight the physical relevance of the NHEB and motivate further investigation into its underlying mechanisms and controllability. Building on these studies, several recent works have explored the NHEB in the presence of spatially nonuniform dissipation by introducing position-dependent loss profiles~\cite{23nonuniformYuce,25imaginarystarkse,25Yuce}. 
However, the role of spatially nonuniform loss in shaping non-Hermitian boundary phenomena remains insufficiently understood. 
In particular, how spatially nonuniform loss influences the NHSE and the NHEB remains an open question.

In this work, we address these issues by systematically studying a one-dimensional non-Hermitian lattice with position-dependent loss. 
By combining magnetic-flux control with spatially varying dissipation, we show that the interplay between these two ingredients gives rise to a range of qualitatively distinct boundary behaviors. 
We find that, even in the absence of the NHSE, corresponding to the zero-flux case, anomalous accumulation of the loss probability persists at the boundary. This boundary accumulation is primarily induced by the loss gradient and exhibits relatively weak spatial localization.
We refer to this behavior as a quasi–non-Hermitian edge burst (quasi-NHEB), distinguishing it from the conventional NHEB associated with the NHSE. 
When magnetic flux is introduced, we find that nonuniform loss and the NHSE interact in a nontrivial manner, leading to a hybrid form of the NHEB whose behavior depends sensitively on system parameters. 
Specifically, the effects of spatially nonuniform loss on the NHSE and on the boundary accumulation of the loss probability depend on the direction of the loss gradient, resulting in either enhanced or suppressed boundary accumulation.
Furthermore, by analyzing both spectral properties and eigenstate distributions, we show that spatially nonuniform loss can significantly modify localization without producing substantial changes in the energy spectrum, indicating a partial decoupling between spectral structure and eigenstate localization.
Finally, we characterize these boundary phenomena through their dependence on the initial position of the quantum walker and their associated bulk–edge scaling relations. 
Our results demonstrate that spatially structured dissipation provides an additional degree of control over non-Hermitian boundary effects, extending beyond the conventional NHSE framework and offering new perspectives for engineered non-Hermitian systems. 

The rest of the paper is organized as follows. In Sec.~\ref{sec:m}, we introduce the non-Hermitian lattice model with magnetic flux and position-dependent loss. In Sec.~\ref{sec:u}, we analyze the flux dependence of the NHSE and the NHEB in the uniform-loss case, establishing a baseline picture. In Sec.~\ref{sec:nu}, we examine the nonuniform-loss case, identify the quasi-NHEB, and discuss the interplay between nonuniform dissipation and the NHSE. In Sec.~\ref{sec:s}, we classify the three types of boundary phenomena and characterize their distinguishing features through initial-state dependence and bulk-edge scaling. Finally, in Sec.~\ref{sec:c}, we summarize our conclusions.

\section{Model}
\label{sec:m}
\begin{figure}[htb]
	\includegraphics[width=0.45\textwidth]{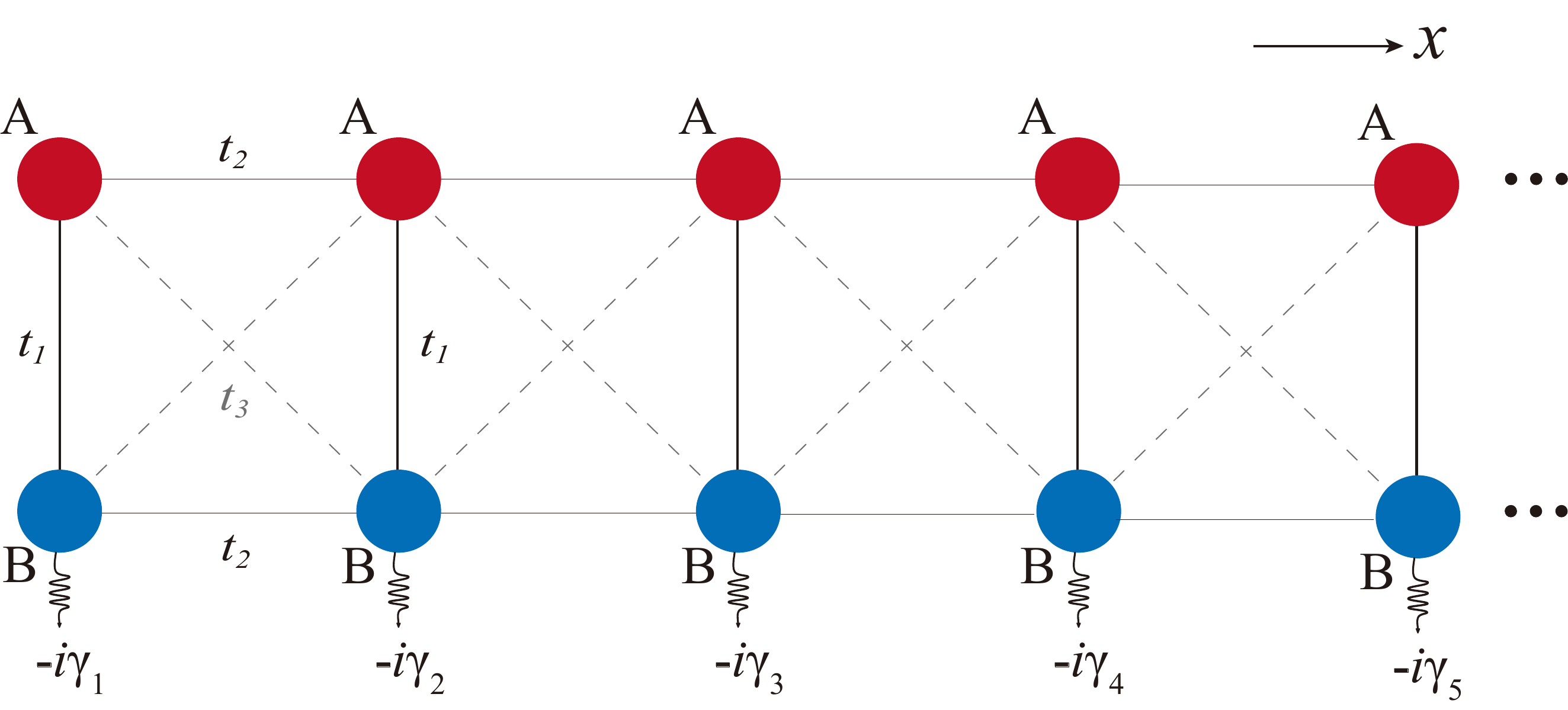}
	\caption{Schematic of a finite tight-binding lattice chain with two sublattices. Loss is introduced only on the B-sublattice (blue sites), with site-dependent strengths $\gamma_x=\{\gamma_1, \gamma_2, \dots, \gamma_L\}$, where $L$ is the total number of unit cells. The system exhibits the NHSE, originating from the interplay between sublattice-selective loss and magnetic flux.
    } 
	\label{Fig.1} 
\end{figure}

We consider a one-dimensional non-Hermitian tight-binding model. The underlying lattice, illustrated in Fig.~\ref{Fig.1}, consists of two coupled sublattice chains, labeled A (red) and B (blue), such that each unit cell contains two sublattice sites. The non-Hermiticity originates from the spatially nonuniform imaginary on-site potential  $-i\gamma_x$ on sublattice B. Here, $t_1$ is the intracell hopping amplitude, whereas $t_2$ and $t_3$ are the intercell nearest-neighbor and next-nearest-neighbor hopping amplitudes, respectively. All hopping amplitudes are taken to be positive real numbers. 
To induce the NHSE, we simultaneously introduce a nonzero loss rate $\gamma_x$ and a finite magnetic flux. In this work, we adopt the Landau gauge for the magnetic field, 
$\mathbf{A}=(0,Bx,0)$. Within this gauge, the magnetic flux is incorporated via the Peierls substitution, which introduces position-dependent phase factors in the hopping amplitudes $t_1$ and $t_3$, namely, $t_1 \to t_{1, x}=t_1 e^{i2\phi x}$ and $t_3 \to t_{3, x} = t_3e^{i 2\phi (x-1/2)}$, where $\phi$ characterizes the magnetic flux threading each triangular plaquette. Accordingly, the time-dependent Schrödinger equation of the resulting model can be written as
\be
 i\frac{d\psi_x^A}{dt}&=&t_{1,x}\psi_x^B+t_2\left(\psi_{x-1}^A-\psi_{x+1}^A\right) \nn \\
 &&+t_{3,x}\psi_{x-1}^B+t_{3,x+1}\psi_{x+1}^B \\
 i\frac{d\psi_x^B}{dt}&=&t_{1,x}^*\psi_x^A+t_2\left(\psi_{x-1}^B-\psi_{x+1}^B\right) \nn \\
 &&+t_{3,x}^*\psi_{x-1}^A+t_{3,x+1}^*\psi_{x+1}^A 
-i\gamma_x\psi_x^B,
\label{Eq.1}
\ee
where $x = 1, 2, \dots, L$, $L$ is the system size, $\psi_{x}^{A}(t)$ and $\psi_{x}^{B}(t)$ are time-dependent complex field amplitudes in the A and B sublattices, respectively.  
We find that both the strength and direction of the NHSE can be continuously tuned by varying the magnetic flux $\phi$. 

To quantify the influence of $\phi$ on the NHSE more clearly, we compute the inverse participation ratio (IPR), defined as:
\begin{equation}
    \operatorname{IPR}^{(i)}=\frac{\sum_{x}\left|\psi_{x}^{i}\right|^{4}}{\left(\sum_{x}\left|\psi_{x}^{i}\right|^{2}\right)^{2}},
\end{equation}
where the superscript $i$ labels the $i$-th eigenstate of the system and $x$ denotes the lattice coordinate. An eigenstate is considered extended when its inverse participation ratio (IPR) scales as $\text{IPR} \sim 1/L$. By contrast, $\text{IPR} \sim 1$ signifies a localized eigenstate confined to a small spatial region, as in the case of Anderson localization~\cite{08ipr}.
However, the IPR characterizes only the localization property of a single eigenstate. To characterize the overall degree of localization of the eigenstates,
we average the IPR over all eigenstates of the system and use the resulting mean IPR (MIPR) as a global indicator of the degree of skin localization~\cite{20MIPR,22MIPR}. Specifically, the MIPR is defined as $\text{MIPR}= {\frac{1}{2L} \textstyle \sum_{i=1}^{2L}\text{IPR}^{(i)}}$.

Having established a quantitative characterization of the NHSE, we next investigate its dynamical consequences through the quantum-walk evolution. We place a quantum walker at the A-sublattice site in the $x_0$-th unit cell at the initial time ($t= 0$), with $x_0$ chosen as close as possible to the center of the lattice. The dynamics of the system are obtained by numerically solving Eq.~(\ref{Eq.1}) under open boundary conditions.
The initial conditions are given by: $\psi_{x}^{A}(t=0)=\delta_{x,x_0}$ and $\psi_{x}^{B}(t=0)=0$. 
During the quantum walk, the walker propagates in discrete steps between the two sublattices and escapes exclusively through the dissipative B sites. As time $t \to \infty$, the walker eventually escapes completely from the lattice. The probability for the quantum walker to escape from a dissipative B site located at position $x$ is given by:
\begin{equation}
P_x=\gamma_x\int_{0}^{\infty}\left| \psi_x^B(t) \right |^2 dt,
\label{Eq.3} 
\end{equation}
which satisfies the normalization condition $\sum_x P_x=1$, with the initial state normalized as $\langle\psi(0)|\psi(0)\rangle=1$.

\begin{figure*}[htbp]
	\includegraphics[width=1\textwidth]{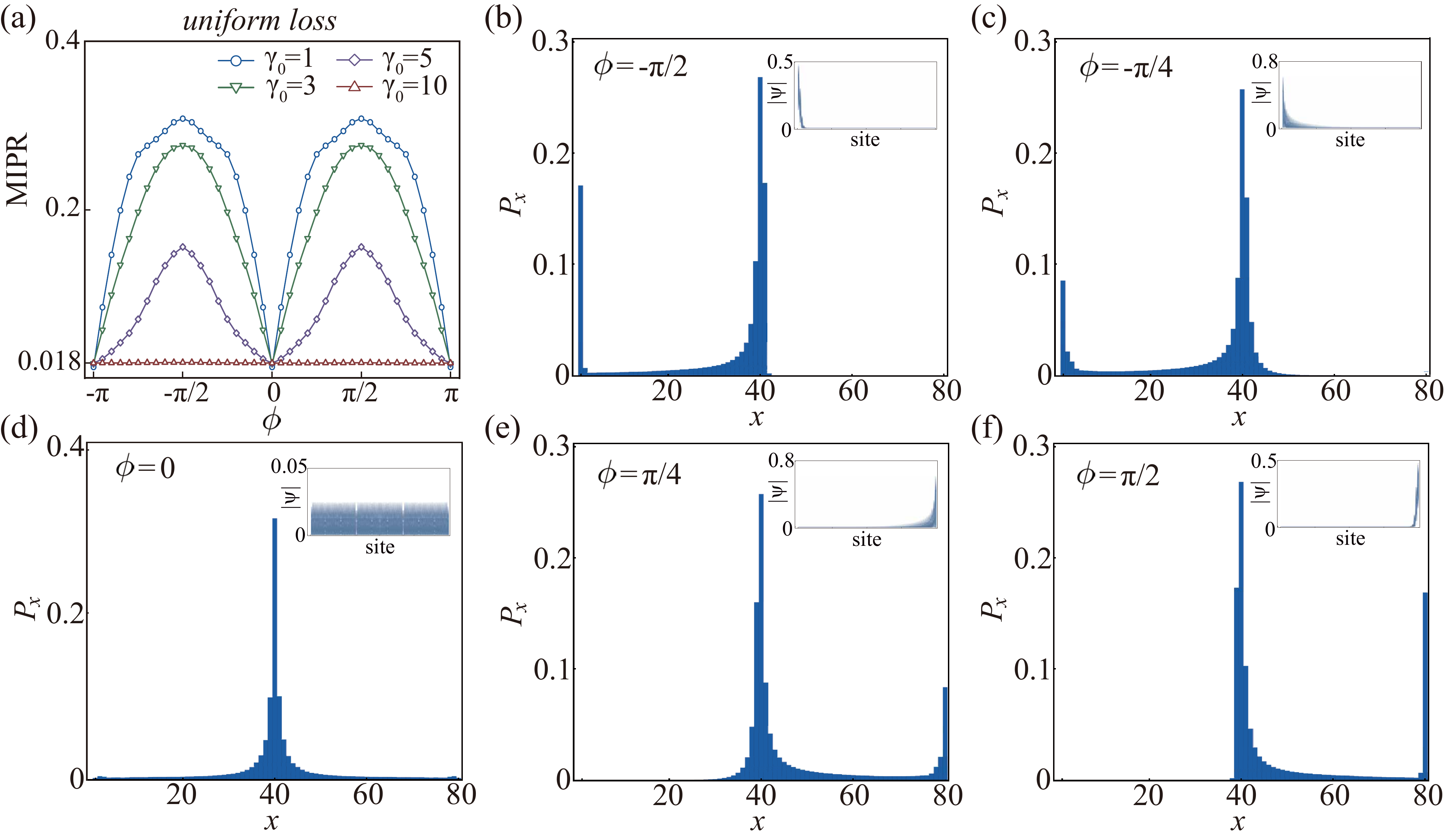}
	\caption{
(a) For the uniform-loss case, mean inverse participation ratio (MIPR) as a function of magnetic flux $\phi$. The horizontal reference $\mathrm{MIPR} \approx 0.018$ corresponds to the absence of the NHSE. In this panel, blue, green, purple, and red curves correspond to uniform loss strengths $\gamma_0 = 1, 2, 5,$ and $10$, respectively.
(b)--(f) Spatial distributions of the loss probability $P_x$ as a function of lattice site $x$ for the uniform loss model, with $\gamma_0=1$. 
Panels (b)--(f) correspond to magnetic flux values $\phi=-\pi/2, -\pi/4, 0, \pi/4, \pi/2$, respectively.  
The initial position of the quantum walker is $x_0=40$.  
Each panel includes an inset showing the probability density of eigenstates under the corresponding flux.  
All panels share system parameters $t_1=0.4$, $t_2=0.25$, $t_3=0.25$, and $L=80$.} 
	\label{Fig.2} 
\end{figure*}

Such dynamical behavior has recently been extensively studied. As demonstrated in Ref.~\cite{21WangZhong}, a distinctive boundary phenomenon, termed the non-Hermitian edge burst (NHEB), can emerge under appropriate parameter conditions. Specifically, in a class of non-Hermitian quantum walks on lossy lattices with open boundary conditions, the distribution of loss probability exhibits an anomalous peak at the system boundary and becomes asymmetric between the left and right sides. It is generally understood to originate from the combined effect of the NHSE and the imaginary gaplessness of the spectrum. However, a systematic understanding of how magnetic flux and spatially nonuniform loss jointly influence edge bursts remains limited. In the following, we systematically investigate their interplay by analyzing the NHSE and the resulting NHEB.

\section{Flux Dependence of the Uniform-Loss Case}
\label{sec:u}
To determine how the variation of the magnetic flux influences the NHSE and the NHEB, we first consider a special and simple case to establish a baseline picture, in which we consider a uniform loss rate with $\gamma_x=\gamma_0$. 

To clearly illustrate the influence of magnetic flux on the NHSE for different values of the uniform loss rate $\gamma_0$, we plot the MIPR as a function of the magnetic flux $\phi$, as shown in Fig.~\ref{Fig.2}(a). Different colors correspond to different values of $\gamma_0$. It should be emphasized that, in the present model, the eigenstates are not perfectly extended, and thus the MIPR does not attain the ideal value of $1/L$. As a result, an MIPR value close to $0.018$ is taken as a practical criterion for the absence of the NHSE throughout this work. This value is not universal but represents the typical MIPR value observed for the present model in the absence of the NHSE over a wide range of parameter settings, and is specific to the system size $L=80$.

For the uniform loss, the MIPR reaches its maximum at $\phi=-\pi/2$ and $\pi/2$, where the skin localization is strongest. In these cases, the eigenstates localize at either the left or right boundary, respectively, which can be attributed to the distinct flux-induced effective chiral motion at $\phi=-\pi/2$ and $\pi/2$. In contrast, at $\phi=0$ and $\pi$, the MIPR is close to 0.018, indicating the absence of the NHSE, as shown in Fig.~\ref{Fig.2}(a) and the inset of Fig.~\ref{Fig.2}(d). Moreover, as $\gamma_0$ increases, the MIPR in the uniform-loss model decreases progressively and eventually approaches the reference value $\mathrm{MIPR}\approx 0.018$ marked in Fig.~\ref{Fig.2}(a), which indicates the absence of the NHSE.

We next focus on how the magnetic flux affects the NHEB, and examine whether its behavior is consistent with the NHSE trend reflected by the MIPR. 
In Fig.~\ref{Fig.2}(b)--(f), we show the NHEB profiles at $\gamma_0 = 1$ for different magnetic fluxes. The insets display the corresponding eigenstate probability density distributions.

As shown in Fig.~\ref{Fig.2}(b), when a magnetic flux $\phi = -\pi/2$ is applied, the loss probability exhibits a pronounced and strongly localized anomalous peak, confined to the unit cell at $x=1$. The loss probability is concentrated on the left side of the initial position $x_0$, consistent with the skin-localization direction of the NHSE. The NHEB reaches its maximum strength, in agreement with the fact that the MIPR is maximized at $\phi=\pm\pi/2$, as shown in Fig.~\ref{Fig.2}(a). For the same hopping parameters, this model is equivalent to that studied in Ref.~\cite{21WangZhong}. When the magnetic flux is increased to $\phi=-\pi/4$, the NHEB peak at the boundary is significantly reduced and becomes broader. This behavior can be attributed to the weakening of the NHSE, as reflected by the reduced MIPR in Fig.~\ref{Fig.2}(a) and the eigenstate probability distribution shown in the inset of Fig.~\ref{Fig.2}(c).

At $\phi=0$, the anomalous loss probability peak at the boundary disappears completely, and the loss probability is concentrated near the initial position $x_0$ and gradually decreases as one moves away from $x_0$, reaching its minimum at the boundary, shown in Fig.~\ref{Fig.2}(d).  
In addition, the NHSE is absent, as illustrated by the inset of Fig.~\ref{Fig.2}(d) and Fig.~\ref{Fig.2}(a). 
Upon further increasing the flux to $\phi=\pi/4$, the NHEB has the same magnitude and width as in the case of $\phi=-\pi/4$, but is localized at the opposite boundary, shown in Fig.~\ref{Fig.2}(e). 
The NHSE at $\phi=\pi/4$ has the same MIPR as that at $\phi=-\pi/4$; however, the direction of the NHSE is reversed, with the localization shifted to the right boundary, as shown in the inset of Fig.~\ref{Fig.2}(e).
Likewise, at $\phi=\pi/2$ the NHEB mirrors that at $\phi=-\pi/2$, except that it is localized at the right boundary.

These results provide a baseline understanding of how the magnetic flux influences the NHSE and the NHEB in this model. For $\phi \in (-\pi,0)$, the NHSE drives eigenstate localization toward the left boundary, and its strength is consistent with the flux dependence of the MIPR shown in Fig.~\ref{Fig.2}(a). Correspondingly, within this flux range, the anomalous peak in the loss probability appears at the left boundary, and its variation with flux is consistent with that of the NHSE. Similarly, for $\phi \in (0,\pi)$, the same behavior and flux dependence persist, but the localization and the associated loss-probability peak are shifted to the right boundary.

\section{Quasi-NHEB under Nonuniform Loss}
\label{sec:nu}
Having established a basic understanding of the magnetic-flux-controlled NHSE in this model, we now focus on the case with nonuniform loss. Without loss of generality, we consider a position-dependent linearly varying loss rate, $\gamma_x=\gamma x$. 

In Ref.~\cite{23nonuniformYuce}, the authors investigated a model with nonuniform loss at a fixed magnetic flux $\phi=-\pi/2$ and reported the occurrence of the NHEB in the apparent absence of the NHSE. Subsequently, Ref.~\cite{25imaginarystarkse} revealed that this system in fact supports a distinct form of NHSE, known as the imaginary-Stark NHSE. Despite these advances, the analysis in both works was largely restricted to $\phi=-\pi/2$, leaving the cases of zero flux and opposite flux unexplored.

Here, we make a full investigation of the interplay between nonuniform loss and magnetic flux. 
We demonstrate that, for $\phi \in (0,\pi)$, nonuniform loss suppresses both the NHSE and the NHEB, despite the presence of finite-area spectral loops under PBCs, which are generally regarded as a hallmark of the NHSE.
Interestingly, we find that, at $\phi=0$, where the NHSE is absent, an edge burst still emerges, we term this NHSE-independent edge burst a quasi–non-Hermitian edge burst (quasi-NHEB).

\begin{figure*}[ht]
	\includegraphics[width=1\textwidth]{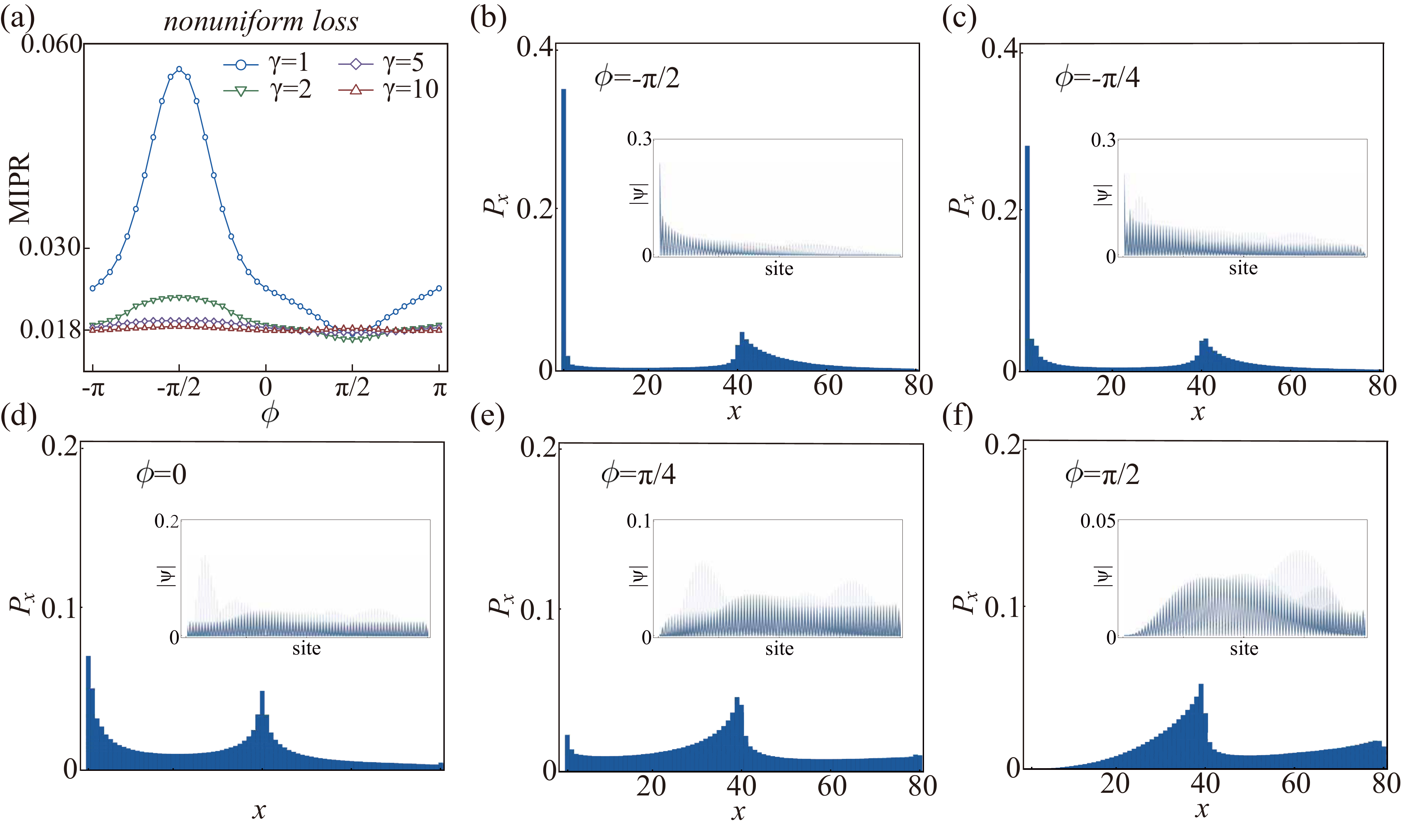}
	\caption{
(a) For the nonuniform loss model, mean inverse participation ratio (MIPR) as a function of magnetic flux $\phi$. The horizontal reference $\mathrm{MIPR} \approx 0.018$ indicates the absence of the NHSE. In this panel, blue, green, purple, and red curves correspond to loss gradients $\gamma = 1, 2, 5,$ and $10$, respectively.
(b)--(f) Spatial distributions of the loss probability $P_x$ as a function of lattice site $x$ for the nonuniform loss model, with $\gamma=1$.  
Panels (b)--(f) correspond to magnetic flux values $\phi=-\pi/2, -\pi/4, 0, \pi/4, \pi/2$, respectively. 
The initial position of the quantum walker is $x_0=40$.  
Each panel contains an inset showing the probability density distributions of eigenstates corresponding to energy levels near the real axis of the complex spectrum under the respective magnetic flux.  
All panels share system parameters $t_1=0.4$, $t_2=0.25$, $t_3=0.25$, and $L=80$.
}
	\label{Fig.3} 
\end{figure*}

We first calculate the MIPR of the system in order to determine the behavior of the NHSE under different parameter conditions. As reported in Ref.~\cite{25imaginarystarkse}, nonuniform loss causes a subset of the eigenstates to become localized on the B sublattice. Accordingly, we evaluate the MIPR as a function of magnetic flux only for the subset of eigenstates associated with the NHSE, as shown in Fig.~\ref{Fig.3}(a). In the range $\phi \in (- \pi, 0)$, the behavior is qualitatively similar to that in Fig.~\ref{Fig.2}(a), except that the MIPR is reduced due to the presence of nonuniform loss, leading to weaker localization strength. Nevertheless, the NHSE still persists in this regime.
In contrast, for the range $\phi \in ( 0, \pi )$, a significantly different behavior emerges. As shown in Fig.~\ref{Fig.3}(a), the MIPR is close to 0.018, indicating that the NHSE is absent throughout this flux range, regardless of the value of $\gamma$. These difference in the MIPR suggest that, within the same system, merely reversing the magnetic flux can qualitatively alter the NHSE. 

Under nonuniform loss rate, we investigate the effects of different magnetic fluxes on the NHEB, as shown in Fig.~\ref{Fig.3}(b)--(f).
Specifically, we begin with the case of magnetic flux $\phi=-\pi/2$. In this setting, we observe a boundary peak in the loss probability whose magnitude is larger than that in the uniform loss case. This peak is almost entirely localized at the boundary site $x=1$. Meanwhile, the loss probability is no longer confined to the localization direction of the NHSE, but instead spreads throughout the entire system, as illustrated in Fig.~\ref{Fig.3}(b). As shown by Fig.~\ref{Fig.3}(a) and the inset of Fig.~\ref{Fig.3}(b), the NHSE reaches its maximum within the nonuniform-loss case, although the MIPR indicates that its localization strength is substantially weaker than that in the uniform loss case. Under the same parameter conditions, our model reduces to that studied in Ref.~\cite{23nonuniformYuce}, and we recover consistent results, thereby confirming the accuracy of our numerical calculations.

Next, we set the magnetic flux to $\phi=-\pi/4$. As shown in Fig.~\ref{Fig.3}(c), the spatial profile of the NHEB undergoes a change similar to that observed in the uniform loss case: the peak of the loss probability on the edge becomes smaller and broader. Correspondingly, the NHSE is also weakened, as illustrated by the MIPR in Fig.~\ref{Fig.3}(a) and the inset of Fig.~\ref{Fig.3}(c).

When the flux is further tuned to $\phi=0$, 
we find that the loss probability still exhibits an enhanced peak localized at the edge, even though both the peak magnitude and the localization are relatively weak. In contrast, both the eigenstate probability distribution in the inset of Fig.~\ref{Fig.3}(d) and the analysis based on the MIPR indicate that the NHSE no longer persists. 
These results demonstrate that the NHEB can emerge even in the absence of the NHSE. To distinguish this NHSE-independent edge burst from the conventional NHEB, we therefore define it as the quasi-NHEB. Specifically, the quasi-NHEB is characterized by weak spatial localization together with an anomalous accumulation of loss probability over a broad boundary region.
In addition, the inset reveals an eigenstate that exhibits a relatively localized probability density near the left side of the chain. This feature is discussed in detail below.

Upon further tuning the magnetic flux to $\phi=\pi/4$ and $\phi=\pi/2$, the system exhibits significantly different behavior. Specifically, at $\phi=\pi/4$, the loss probability is distributed throughout the entire system, with only a very weak anomalous peak remaining at the left boundary, as shown in Fig.~\ref{Fig.3}(e). 
This behavior is markedly different from that observed in the case of uniform dissipation. Under uniform dissipation, when the magnetic flux is varied from $\phi=-\pi/4$ to $\phi=\pi/4$, the NHSE retains the same strength while reversing its localization direction. Therefore, in the nonuniform loss case, one may expect a weak edge burst to emerge at the right boundary for $\phi=\pi/4$. Surprisingly, however, no NHEB is observed at the right boundary in this regime as illustrated in Fig.~\ref{Fig.3}(e). Moreover, the NHSE disappears completely, as evidenced by Fig.~\ref{Fig.3}(a) and the eigenstate probability distribution shown in the inset of Fig.~\ref{Fig.3}(e).
Similarly, for $\phi=\pi/2$, the loss-probability distribution shown in Fig.~\ref{Fig.3}(f) exhibits no boundary peak at all. Although the NHSE would be expected to be strongest at $\phi=\pi/2$, Fig.~\ref{Fig.3}(a) and the inset of Fig.~\ref{Fig.3}(f) show that no NHSE exists in this regime; indeed, the corresponding MIPR is even smaller than that at $\phi=\pi/4$. 

\begin{figure}[h]
	\includegraphics[width=0.45\textwidth]{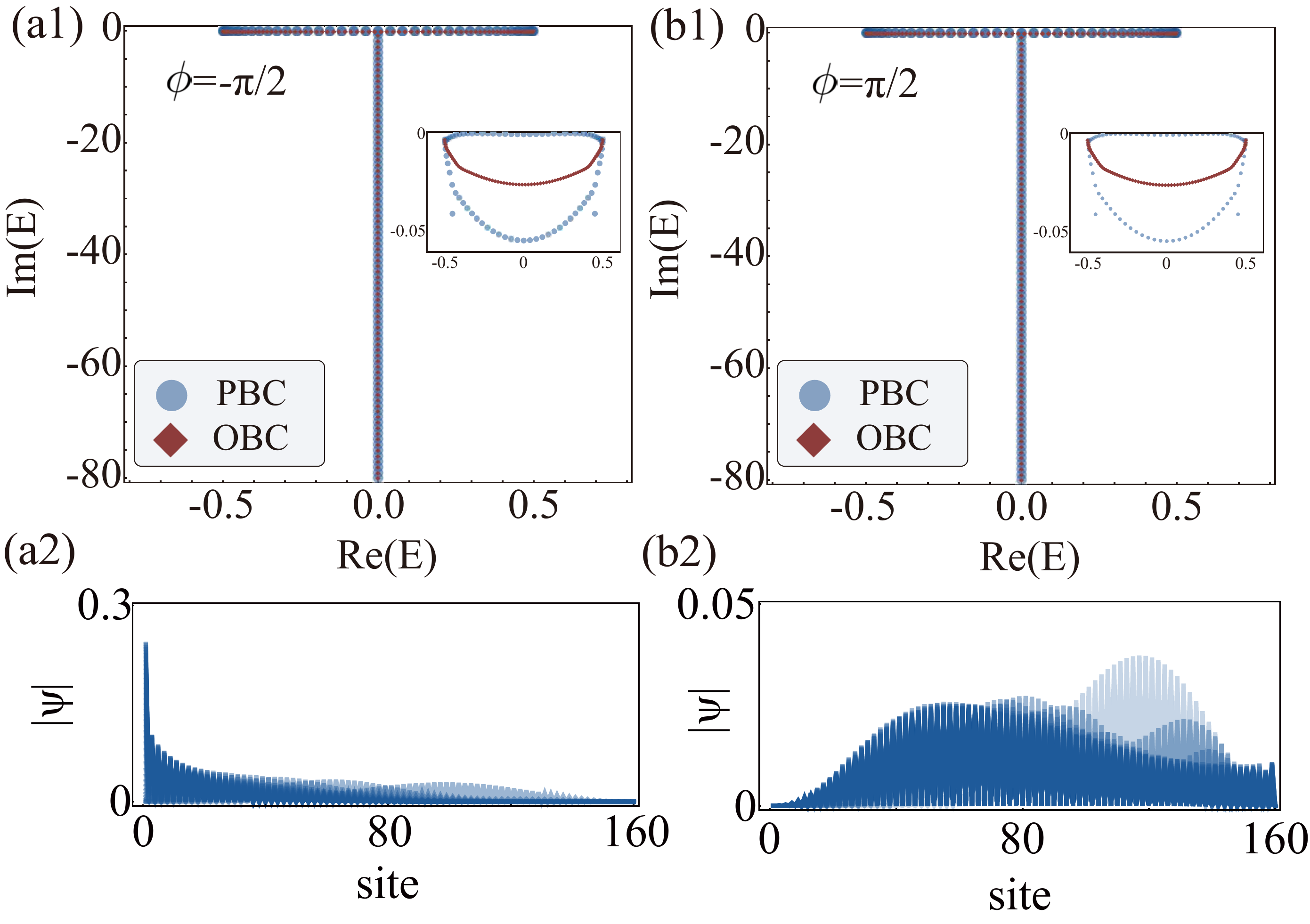}
	\caption{
Complex energy spectra of the nonuniform loss model for (a1) $\phi=-\pi/2$ and (b1) $\phi=\pi/2$.  
Red diamonds and blue circles denote spectra under open and periodic boundary conditions, respectively. The insets in panels (a1) and (b1) show enlarged views of the complex energy spectrum in the vicinity of the real axis. 
(a2) and (b2) show the eigenstate probability density distributions corresponding to energy levels near the real axis for the $\phi=-\pi/2$ and $\phi=\pi/2$, respectively.  
System parameters are $\gamma=1$, $t_1=0.4$, $t_2=0.25$, $t_3=0.25$, and $L=80$.} 
	\label{Fig.4} 
\end{figure}

To gain further insight into the origin of the pronounced difference in the loss probability between $\phi=-\pi/2$ and $\phi=\pi/2$, we plot the corresponding complex energy spectra in Fig.~\ref{Fig.4}. In the nonuniform loss case, the spectrum exhibits a characteristic ``T''-shaped structure. The portion of the spectrum located near the real axis is the part relevant to the NHSE. In contrast, the vertical branch along the imaginary axis shows identical spectra under both PBC and OBC. The eigenstates associated with this branch exhibit strong localization, with each eigenstate localized around an individual B-sublattice within the bulk region, and are therefore unrelated to the NHSE~\cite{25imaginarystarkse}.

Focusing on the energy spectra, we find that the complex spectra corresponding to $\phi=-\pi/2$ and $\phi=\pi/2$ exhibit nearly identical overall structures. Moreover, upon magnifying the portion of the spectrum near the real axis, it can be seen that the PBC spectrum encloses the OBC spectrum. According to the conventional spectral criterion for the NHSE~\cite{22PBCacceleration,22PBCarea=nhse,201dPBCNHSE}, this enclosing relation indicates that the system should exhibit the NHSE for both flux values under the corresponding parameter conditions. If we relied only on this spectral criterion, together with the correspondence between magnetic flux and the NHSE discussed previously for the uniform loss case, we would be led to infer that the NHSE in the two cases should exhibit similar strength but opposite localization directions.

This conclusion, however, appears inconsistent with the results obtained from the MIPR and the NHEB. To resolve this discrepancy, we enlarge the inset panels in Fig.~\ref{Fig.3}(b) and Fig.~\ref{Fig.3}(f), and the enlarged distributions are presented in Fig.~\ref{Fig.4}(a2) and Fig.~\ref{Fig.4}(b2). The eigenstate density profiles for $\phi=-\pi/2$ and $\phi=\pi/2$, corresponding to the eigenvalues near the real axis of the complex spectra in Fig.~\ref{Fig.4}(a1) and Fig.~\ref{Fig.4}(b1), are significantly different: the former exhibits clear NHSE-induced localization, whereas the latter indicates the absence of the NHSE. These results indicate that the NHSE is indeed absent at $\phi=\pi/2$. We further reveal that nonuniform loss can suppress eigenstate localization even when the spectral structure suggests the presence of the NHSE, thereby affecting both the emergence and the strength of the NHEB.

To quantify the effect of nonuniform loss on the NHSE, we evaluate the MIPR at $\phi=\pm\pi/2$. We find that at $\phi=\pi/2$, the MIPR value (0.017) differs markedly from that at $\phi=-\pi/2$ (0.056), and is close to $0.018$, indicating the absence of the NHSE. To more directly illustrate the magnitude of this influence, we further introduce a relative change ratio, defined as:
\begin{equation}
   \frac{\mathrm{MIPR}^{(1)}-\mathrm{MIPR}^{(2)}}{\mathrm{MIPR}^{(2)}},
\end{equation}
where superscripts $(1)$ and $(2)$ correspond to the cases $\phi=-\pi/2$ and $\phi=\pi/2$, respectively. This relative variation reaches as high as $229.4\%$, suggesting that the spatial distribution of nonuniform loss can strongly influence the strength of the NHSE. Together with the flux $\phi=0$ case discussed above, this result further supports the conclusion that nonuniform loss can substantially influence the localization properties of the eigenstates, and affect both the emergence and the strength of the NHEB.

\begin{figure}[h]
	\includegraphics[width=0.45\textwidth]{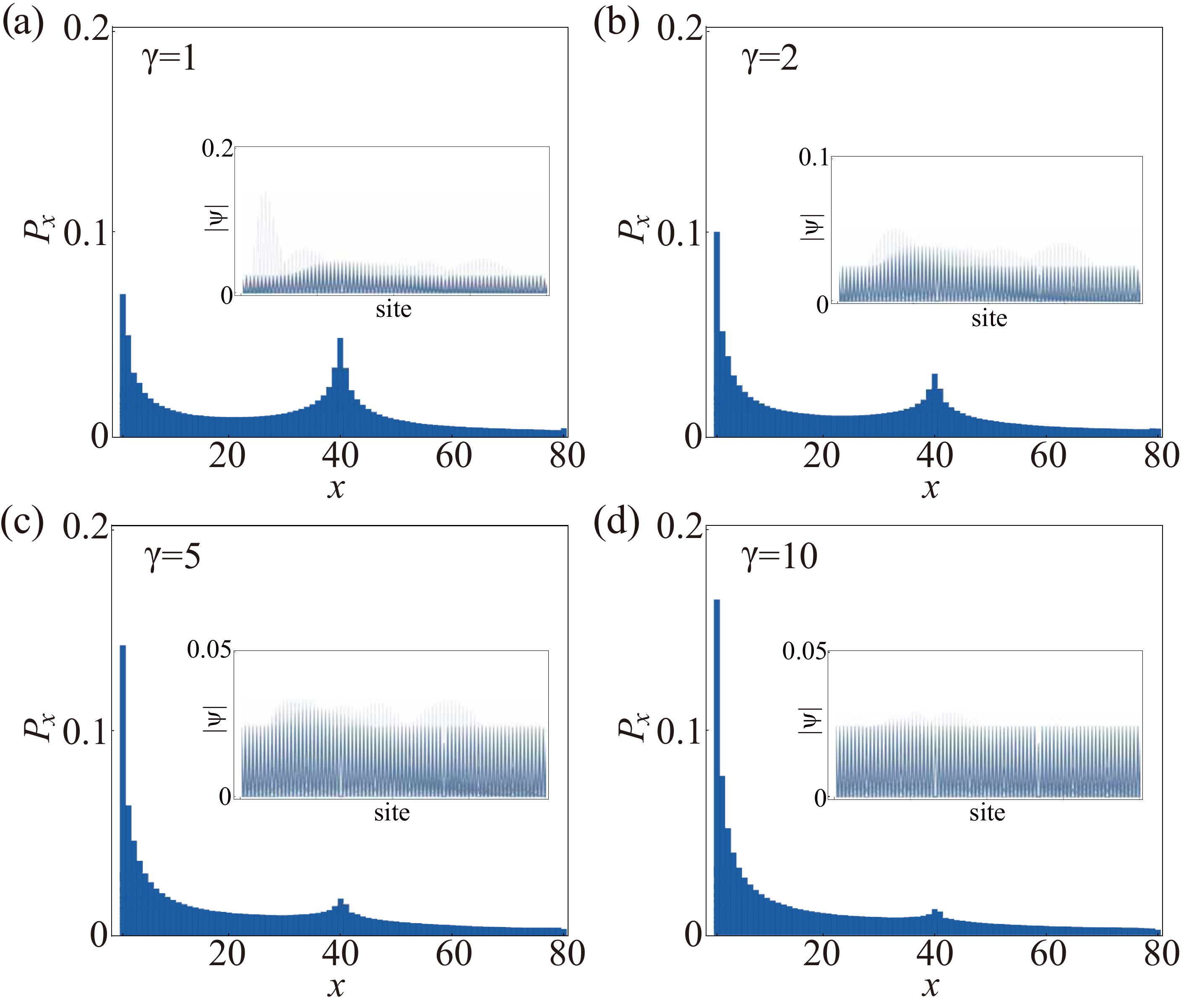}
	\caption{
Spatial distributions of the loss probability $P_x$ for the nonuniform loss model at zero magnetic flux ($\phi=0$).  
Panels (a)--(d) correspond to $\gamma=1,2,5,10$, respectively.  
All panels include insets showing eigenstate probability densities corresponding to energy levels near the real axis of the complex spectrum.  
The initial position of the quantum walker is $x_0=40$, and system parameters are $t_1=0.4$, $t_2=0.25$, $t_3=0.25$, and $L=80$.}
	\label{Fig.5} 
\end{figure}

To gain deeper insight into the quasi-NHEB and to examine whether it can be induced and controlled solely by the nonuniform loss gradient $\gamma$, we set the magnetic flux to $\phi=0$ and vary $\gamma$ to investigate the evolution of the boundary loss peak, as illustrated in Fig.~\ref{Fig.5}. The insets of Fig.~\ref{Fig.5} show the probability density distributions of the eigenstates associated with the eigenvalues near the real axis of the complex spectrum. 
In Fig.~\ref{Fig.5}(a), we presents the quasi-NHEB for $\gamma=1$, which corresponds to Fig.~\ref{Fig.3}(d). When $\gamma$ is increased to $\gamma=2$, the boundary peak is clearly enhanced, while the loss probability near the initial position $x_0$ is reduced. At the same time, the inset shows that the eigenstates become increasingly similar to extended states, consistent with the MIPR shown in Fig.~\ref{Fig.3}(a). The same trend persists for $\gamma=5$ and $10$. For these larger values of $\gamma$, the boundary loss-probability peak continues to increase, whereas the loss probability near the $x_0$ keeps decreasing, as shown in Figs.~\ref{Fig.5}(c) and \ref{Fig.5}(d). Notably, these results indicate that, for the quasi-NHEB, the anomalous accumulation peak at the boundary is positively correlated with $\gamma$, whereas the peak at the $x_0$ is negatively correlated with $\gamma$. In particular, at $\gamma=10$, $P_{\mathrm{edge}}$ becomes much larger than $P_{x_0}$, indicating that the quantum walker is more likely to escape near the boundary than near the initial site $x_0$. This indicates that, within this parameter regime, although the magnitude of the boundary peak remains moderate, the quasi-NHEB is already pronounced.

Based on the numerical results above, we conclude that the quasi-NHEB differs from the conventional NHEB in several respects. First, in terms of its physical origin, the quasi-NHEB is not associated with the NHSE, but is instead governed by $\gamma$. In the nonuniform loss case without magnetic flux, $\gamma$ directly determines the strength of the quasi-NHEB and also affects the loss probability near the $x_0$. By contrast, in the uniform loss case, $\gamma_0$ influences the NHEB only indirectly through its effect on the NHSE. It is worth noting that the energy spectrum of the nonuniform loss case still satisfies one of the conditions for the emergence of the conventional NHEB, namely, the imaginary gap closing. Therefore, we specifically mean that the effect of nonuniform loss is manifested without being influenced by the NHSE.
In addition, as shown previously in Fig.~\ref{Fig.3}, when the NHSE is induced by magnetic flux, the effect of nonuniform loss can arise independently and may, for certain parameters, suppress the NHEB governed by the NHSE.

Second, from a phenomenological point of view, the quasi-NHEB differs from the conventional NHEB in its spatial profile. The conventional NHEB gives rise to a pronounced anomalous loss peak at the boundary site and, due to the NHSE, the loss probability is nearly absent on the side opposite to the skin-localization direction. By contrast, the quasi-NHEB manifests as an anomalous accumulation of loss probability over several unit cells near the boundary, with the largest peak still located at the boundary, while finite loss probability remains distributed throughout the entire system. 

This anomalous accumulation of loss probability at the boundary, induced solely by nonuniform dissipation, is distinct from the conventional NHEB, whose emergence and strength are governed by the NHSE. We refer to the former as the quasi-NHEB. In the presence of both flux-induced NHSE and nonuniform loss, the resulting edge burst originates from the interplay between these two mechanisms and is therefore referred to as the hybrid NHEB, as illustrated in Figs.~\ref{Fig.3}(b) and \ref{Fig.3}(c).

Finally, we note that the inset of Fig.~\ref{Fig.5}(a) reveals a relatively localized eigenstate near the left edge of the chain, suggesting a possible connection to the emergence of the quasi-NHEB. However, with increasing loss gradient $\gamma$, this eigenstate gradually loses its localization and merges into the bulk, whereas the quasi-NHEB becomes progressively stronger. At $\gamma$=10, all eigenstates are nearly extended, while the quasi-NHEB reaches its maximum strength. These observations rule out this localized eigenstate as the origin of the quasi-NHEB.

This behavior may be understood from the role played by the imaginary on-site potential associated with the nonuniform loss. Besides providing particle loss, the imaginary on-site potential also suppresses intercell hopping. A similar mechanism was recently reported in Ref.~\cite{26impurity}, where a sufficiently large imaginary on-site potential was shown to generate an effective boundary inside the bulk, acting as an effective potential barrier. 
In the present model, the imaginary on-site potential increases linearly with position owing to the linearly varying loss profile, thereby giving rise to an effective potential barrier with a linearly increasing height. Consequently, preferential hopping toward the B sublattice of unit cells with smaller local loss occurs, followed by particle escape from the system. As the loss gradient $\gamma$ increases, the effective imaginary potential becomes steeper in space, leading to a more pronounced accumulation of the loss probability near the boundary and hence a stronger quasi-NHEB.

\section{Bulk-edge scaling and initial-state dependence of the NHEB}
\label{sec:s}
\begin{figure}[h]
	\includegraphics[width=0.45\textwidth]{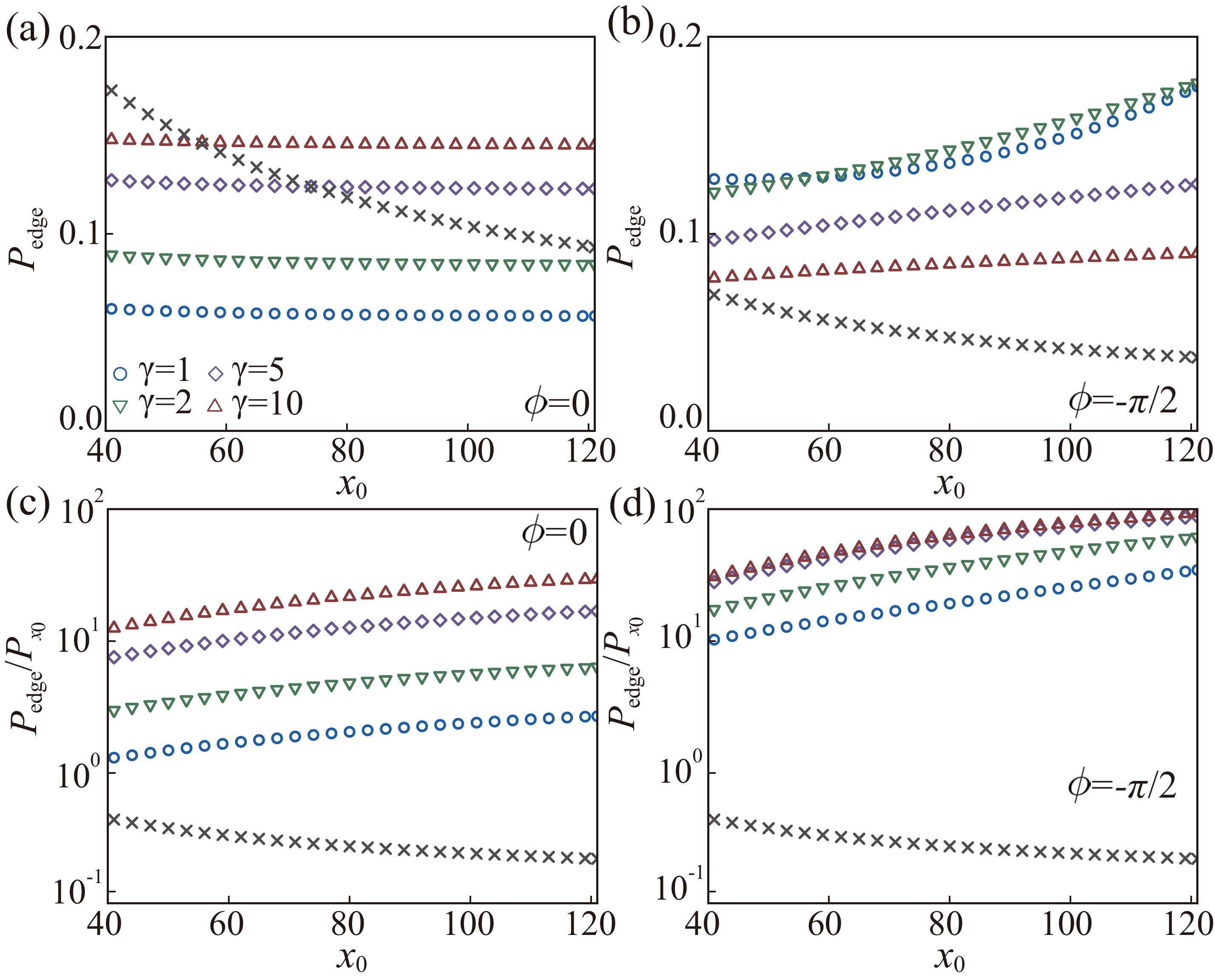}
	\caption{
Dependence of boundary loss probability on the initial position $x_0$ for the nonuniform loss case. 
(a) $P_{\mathrm{edge}}$ as a function of $x_0$ at zero magnetic flux ($\phi = 0$). 
(b) Same as (a), but for $\phi = -\pi/2$. 
(c) Logarithmic plot of the relative loss probability $P_{\mathrm{edge}}/P_{x_0}$ versus $x_0$ at $\phi = 0$. 
(d) Same as (c), but for $\phi = -\pi/2$. 
In all panels, blue, green, purple, and red symbols correspond to loss gradients $\gamma = 1, 2, 5,$ and $10$, respectively.  
The symbol ``$\times$'' represents the conventional NHEB in the uniform-loss model at $\phi = -\pi/2$, shown in all panels solely as a reference for comparison.
The initial position $x_0$ is chosen deep in the bulk to minimize boundary effects. 
Other parameters are $t_1 = 0.4$, $t_2 = 0.25$, $t_3 = 0.25$, and system size $L = 150$.} 
	\label{Fig.6} 
\end{figure}

To more clearly elucidate the distinct character of the quasi-NHEB, we classify the NHEB-related phenomena discussed in this work into three representative cases: (i) the quasi-NHEB induced solely by nonuniform loss in the absence of magnetic flux; (ii) the hybrid NHEB arising from the interplay between the NHSE and nonuniform loss in the presence of a magnetic flux $\phi=-\pi/2$; and (iii) the conventional NHEB in the uniform-loss model at $\phi=-\pi/2$, which serves as a reference case. The corresponding results are summarized in Fig.~\ref{Fig.6}.

As shown in Fig.~\ref{Fig.6}, we plot the boundary loss probability $P_{\mathrm{edge}}$ and the relative loss probability $P_{\mathrm{edge}}/P_{x_0}$ as functions of the initial position $x_0$ of the quantum walker. To focus on bulk behavior and minimize boundary-induced effects, we consider a larger system size and choose $x_0$ deep in the bulk.

As illustrated in Fig.~\ref{Fig.6}(a), for the quasi-NHEB [case (i)], $P_{\mathrm{edge}}$ remains nearly independent of the initial position $x_0$ for all considered values of $\gamma = 1, 2, 5,$ and $10$. In addition, $P_{\mathrm{edge}}$ increases with $\gamma$. By contrast, for the conventional NHEB (denoted by ``$\times$''), $P_{\mathrm{edge}}$ exhibits a power-law decay as the initial position $x_0$ moves away from the left boundary~\cite{21WangZhong}. These results indicate that, except when $x_0$ is very close to the boundary, the quasi-NHEB is essentially insensitive to $x_0$ and is governed primarily by $\gamma$.

We further compare the hybrid NHEB [case (ii)] with the conventional NHEB, as shown in Fig.~\ref{Fig.6}(b). For the hybrid NHEB, the boundary loss probability $P_{\mathrm{edge}}$ increases as the distance between the initial position $x_0$ and the boundary increases. Moreover, the growth rate decreases with increasing $\gamma$. This behavior reflects the nontrivial interplay between the NHSE and nonuniform loss in the presence of magnetic flux.

To further clarify the relative scaling between the boundary loss probability and the initial-site loss probability, we plot the logarithmic dependence of $P_{\mathrm{edge}}/P_{x_0}$ on $x_0$ in Figs.~\ref{Fig.6}(c) and \ref{Fig.6}(d). For the quasi-NHEB [Fig.~\ref{Fig.6}(c)], $P_{\mathrm{edge}}/P_{x_0}$ increases as $x_0$ moves away from the boundary, and the growth rate is positively correlated with $\gamma$. Since $P_{\mathrm{edge}}$ remains nearly constant [see Fig.~\ref{Fig.6}(a)], this implies that $P_{x_0}$ decreases with increasing distance from the boundary, and that the decrease in $P_{x_0}$ becomes more pronounced as $\gamma$ increases. In contrast, for the conventional NHEB in the uniform loss model, $P_{\mathrm{edge}}/P_{x_0}$ continues to decrease as $x_0$ is moved away from the boundary, suggesting that $P_{x_0}$ varies more weakly with $x_0$ than $P_{\mathrm{edge}}$.

For the hybrid NHEB, the overall trend is similar to that of the quasi-NHEB, namely, $P_{\mathrm{edge}}/P_{x_0}$ increases as $x_0$ is shifted away from the boundary region, as shown in Fig.~\ref{Fig.6}(d). However, unlike the behavior of $P_{\mathrm{edge}}$ in Fig.~\ref{Fig.6}(b), the magnitude of $P_{\mathrm{edge}}/P_{x_0}$ is positively correlated with $\gamma$.

From these results, we draw the following conclusions. The quasi-NHEB induced by nonuniform loss is qualitatively distinct from both the hybrid NHEB and the conventional NHEB: it is essentially insensitive to the initial position $x_0$, while its magnitude increases with $\gamma$. 
Furthermore, the nontrivial interplay between nonuniform loss and the NHSE gives rise to the ``imaginary Stark skin effect''~\cite{25imaginarystarkse}, resulting in the hybrid NHEB with rich and intricate dynamical behavior. In addition, the conventional NHEB is solely induced and controlled by the NHSE arising from nonreciprocity.


\section{Conclusion}
\label{sec:c}
In this work, we have systematically investigated the role of spatially nonuniform loss in non-Hermitian quantum walks on lossy lattices with open boundaries. By introducing a magnetic-flux-controlled non-Hermitian model, we demonstrated that the interplay between magnetic flux and dissipation provides a controllable platform for tuning both the strength and direction of the NHSE.

Our results show that nonuniform loss qualitatively modifies boundary phenomena associated with non-Hermitian systems. In the absence of magnetic flux, where the NHSE is absent, we identified a boundary accumulation of loss probability that persists and grows with the loss gradient. Owing to its weaker localization and extended spatial profile, we refer to this phenomenon as a quasi-NHEB, which suggests that boundary-localized dissipation anomalies can arise independently of the NHSE.

Upon introducing a magnetic flux, nonuniform loss and the NHSE exhibit a nontrivial interplay. Depending on the magnetic flux, this interplay can either enhance or suppress boundary accumulation, leading to a hybrid form of the NHEB with behavior distinct from that of the conventional case. We further showed that these different regimes can be clearly distinguished through their dependence on the initial position of the quantum walker and their corresponding bulk–edge scaling relations.

Furthermore, by analyzing both the energy spectra and eigenstate distributions, we demonstrated that spatially nonuniform loss can significantly alter the localization properties of eigenstates without necessarily producing pronounced changes in the spectral structure. This observation indicates that spectral criteria alone may not fully capture the emergence of boundary phenomena in non-Hermitian systems with inhomogeneous dissipation.

More broadly, our results suggest that spatially structured dissipation offers an additional route for controlling non-Hermitian boundary effects beyond the conventional NHSE framework. These findings provide a complementary perspective on the origin of boundary-localized phenomena and may have implications for future studies of engineered dissipation in synthetic quantum and photonic systems. Possible extensions of the present work include higher-dimensional lattices, where the interplay between magnetic flux and nonuniform loss may lead to even richer boundary phenomena, as well as experimental realizations in photonic waveguide arrays and electric circuit networks, where position-dependent loss can be implemented with high precision.


\bibliography{references}
\bibliographystyle{apsrev4-2}
\end{document}